# *In situ* Study of Phase Transitions in $La_2NiO_{4+\delta}$ ($0 \leq \delta \leq 0.14$) using Raman Spectroscopy

Adeel Riaz [a,b*], Alexander Stangl [a], Mónica Burriel [a], Michel Mermoux [b*]

[a] Univ. Grenoble Alpes, CNRS, Grenoble INP, LMGP, Grenoble 38000, France.
[b] Univ. Grenoble Alpes, Univ. Savoie Mont Blanc, CNRS, Grenoble INP, LEPMI, Grenoble 38000, France

* E-mail: adeel.riaz@grenoble-inp.fr; michel.mermoux@grenoble-inp.fr

**Abstract:** $La_2NiO_{4+\delta}$ has attracted increasing interest in recent years, both as oxygen electrode in solid oxide fuel cells and electrolysers due to its high electrochemical activity at intermediate to high temperatures, and as key component of memristive devices for neuromorphic computing, owing to its variable oxygen stoichiometry. The integration of $La_2NiO_{4+\delta}$ into devices operating at different temperatures and oxygen partial pressures requires knowledge of the effects of hyper-stoichiometry ($\delta$) on its crystalline structure. $La_2NiO_{4+\delta}$ is known to accommodate oxygen at interstitial sites allowing for large $\delta$ values, up to ~ 0.16. In addition, the O-doping - temperature phase diagram is known to be complex, exhibiting several phase transitions with increasing $\delta$. Herein, we use Raman spectroscopy to monitor the effects of O-doping in the phase diagram and the various structures it contains. Throughout this work, we studied this material in its usual ceramic form, as well as in the form of thin films. Results are discussed in terms of phase transitions, chemical expansion, and some of the possible consequences of the low mean grain size inherent to such thin films.

**Keywords**: Lanthanum nickelate, Ruddlesden-Popper phases, in situ Raman spectroscopy, ceramics

## 1. Introduction

In recent years, one of the main challenges in the field of reversible Solid Oxide Cells (rSOCs) materials has been devoted to lower the operating temperature (T), typically well below 800 °C [1–5]. For such a purpose, potential candidates for cathodic materials must fulfill several requirements, namely (i) high electronic and ionic conductivities, (ii) high catalytic activity for oxygen reduction reaction (ORR) and oxygen evolution reaction (OER) as well as (iii) high stability under working conditions. In this context, rare-earth nickelates of formula $Ln_2NiO_{4+\delta}$ (Ln = La, Nd, Pr), have been widely studied as potential mixed ionic-electronic conducting (MIEC) oxygen electrode materials [6–9]. Throughout this paper, we will focus on lanthanum nickelate, $La_2NiO_{4+\delta}$ (L2NO4). In particular, it has recently been shown that tailored, architecturally designed in the form of thin films, L2NO4 oxygen electrodes can effectively improve the charge transfer kinetics by increasing the specific surface area [10]. This specific thin film architecture may be considered a first step to enable operating temperatures around 500 °C.

Rare-earth nickelates, and more generally their isostructural cuprate counterparts, were largely studied in the 90's for their low temperature electronic transport and magnetic properties, and most importantly for their superconductivity at high temperatures [11–16]. These oxides belong to the series of the Ruddlesden-Popper (R-P) family ($Ln_{n+1}NiO_{3n+1}$), being the n = 1 member of the family [17]. Interestingly, rare-earth nickelates are known to accommodate interstitial oxygen, charge compensated by electron holes localized on the transition metal cation, leading to the observed mixed ionic and electronic conductivity. The available range of interstitial oxygen content or oxygen hyper-stoichiometry ($\delta$) is known to depend on the size of the Ln cation radii and to be influenced by the synthesis and/or annealing conditions [18]. Concerning L2NO4, reported $\delta$ values usually lie in the 0 - 0.16 range, but it was shown that specific electrochemical processes can increase interstitial oxygen content up to $\delta \approx 0.3$ [19].

The structural and vibrational properties of L2NO4 as a function of $\delta$ can be studied using Raman Spectroscopy (RS). It is an inelastic light scattering technique, mostly described as a vibrational spectroscopy method. However, this description can be somewhat restrictive, as in this excitation frequency range, incident photons essentially probe electrons, while vibration modes just modulate the sample's electronic susceptibility and, hence, its optical properties. Nevertheless, we will adhere to this conventional description in the following discussion. Clearly, when analyzing (single) crystals, X-ray diffraction (XRD) and RS complement each other, although not all space groups (SG) can be unambiguously identified by both methods [20], as each technique follows its own "selection rules". One of the key differences between the two methods is that XRD probes mainly cationic sub-lattices, while RS is usually sensitive to oxygen-dominated vibrational modes. This is especially true in the high frequency region of the vibrational spectra of oxide materials, where oxygen makes a dominant contribution to the vibrational density of state (VDOS). The effects of disorder in the anionic sub-lattice can, therefore, be assessed, if necessary, although the result will depend on the material and the extent of the deviations from stoichiometry. Both methods are also sensitive to strain/stress effects, through the observation of Raman line (or XRD peaks) shifts and/or splitting, depending on the crystal symmetry and on the nature/symmetry of the strain, isotropic or anisotropic. According to the quasi-harmonic approximation, the strain dependence of the phonon wavenumbers may be understood introducing the so-called Grüneisen parameters [21]. In its microscopic formulation, the Grüneisen parameter $\gamma$ is a measure of how a specific phonon frequency $\omega$ is altered under a small change in the geometry of the crystallographic unit cell. Although this approach may prove insufficient, particularly due to the isotropic assumption it is based on, which regards $\gamma$ as a scalar quantity, line shifts are still anticipated when the unit cell undergoes contraction or expansion. A simple rule applies: "the closer the atoms, the larger the spring constant values". In most instances, this rule holds true and, at least qualitatively, explains the observed line shifts with reliable orders of magnitude. What is more, RS can be sensitive to disorder in the anionic sub-lattice, through the detection of forbidden modes that violate selection rules, or even broad, continuous signals that may reflect VDOS-like features. Finally, oxygen doping can also modify possible resonant scattering conditions, whether from disorder (simple smoothing of resonance conditions), or from chemical expansion considerations (variation in lattice parameters). For such situations, the analysis should be conducted for different excitation wavelengths. In particular, deviations from stoichiometry can lead to dramatic changes in the shape of the Raman spectra of specific oxides, as those crystallizing in the fluorite structure, see for example ref [22] that concerns cerium oxide. In these specific cases, it is not uncommon to observe forbidden, infrared symmetry-allowed longitudinal optical (LO) modes in the Raman spectra. In most cases, a Fröhlich electron-phonon [23] interaction is invoked to analyze/understand the presence of these modes in the Raman spectra.

As previously mentioned, in terms of applications, L2NO4 has gained interest in recent years as an oxygen electrode in reversible solid oxide cells (rSOC) due to its high electrochemical activity at intermediate to high temperatures [6,10,24] and for memristive devices for neuromorphic computing [25,26]. From a more academic perspective, L2NO4 seems quite amenable to study for four different reasons: i) L2NO4 can accommodate oxygen at interstitial sites giving large over-stoichiometry $\delta$ values, up to ~ 0.16 or even higher, ii) the O-doping phase diagram ($T - \delta$) of L2NO4 is known to be complex (and most likely sample-dependent, as indicated by the existing literature), iii) Raman data on L2NO4 are limited [15,27,28], with no clear consensus among the few available datasets, iv) L2NO4 is effectively intended for use as a thin layer in various functional devices. The rationale behind this study was to use RS, in complement to XRD, to explore the high temperature behavior of L2NO4 in the form of dense pellets and thin films to probe the O-doping phase diagram.

## 1.1. $La_2NiO_{4+\delta}$ structure and phase diagram

The layered structure of L2NO4, which belongs to the $K_2NiF_4$-type structure, is built of perovskite-type $LaNiO_3$ and rock-salt-type LaO layers alternating along the crystallographic c-axis. Because of the small size of the La ion, the La–O bonds are in extension while the Ni–O bonds are compressed. These distortions relax in different ways by inserting oxygen in the La–O layer. As a result, the O-doping phase diagram ($T - \delta$) is rather complex, and presents a succession of ordered domains and phase separation regions. The O-doping is not isovalent, meaning that holes are created for charge compensation, accompanied by a formal change of the Ni oxidation state. The hole concentration is usually approximated as $2\delta$, which is likely a rough estimate, particularly in the case of thin films. The $T - \delta$ phase diagram of bulk L2NO4 has been studied over the years by several groups with different techniques, in particular X-ray and neutron diffraction [19,29–35]. Based on the literature, not all authors provided consistent boundaries or the same sequence of phases. Some of the key features are represented by the schematic phase diagram proposed by Odier *et al.* [35], which is reproduced in Figure 1, in particular the sequence of monophasic and biphasic domains. Synthesis methods have been suggested as the key factor influencing the determination of boundaries between the different phases formed [29,30,32]. Crystallite size variations may represent another parameter influencing phase transitions. To the best of our knowledge, this factor has not been previously considered.

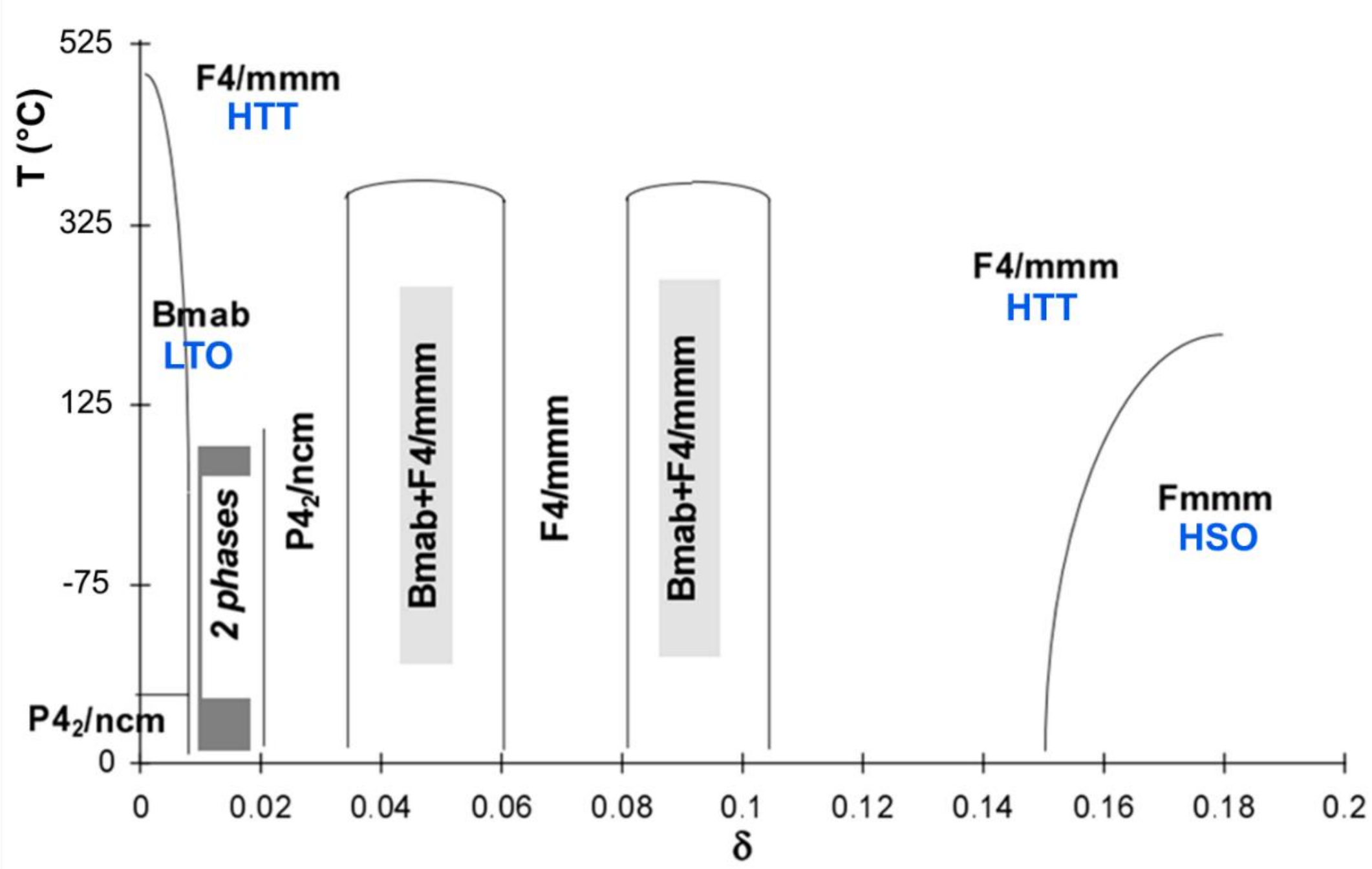

Figure 1: Schematic O-doping phase diagram (T – δ) for L2NO4, based on the phase diagram from [35]; LTO: low temperature orthorhombic, HTT: high temperature tetragonal, HSO, high stoichiometry orthorhombic.

To describe this O-doping phase diagram, it is common to first consider the ideal phase with highest symmetry. This ideal L2NO4 structure, commonly referred to as the High Temperature Tetragonal (HTT) phase, adopts a tetragonal structure with space-group (SG) I4/mmm or $D_{4h}^{17}$. For the exact stoichiometric composition (δ = 0) this phase is observed at elevated temperatures, typically above ~ 500 °C. This ideal tetragonal structure corresponds to an almost perfect matching of the bond lengths of the perovskite and rock-salt layers. However, as a function of temperature and δ, the crystal cell often presents a lower symmetry than the ideal $K_2NiF_4$ tetragonal parent structure. In most studies, focusing on temperatures above ambient and for increasing δ values, L2NO4 undergoes sequential phase transitions from orthorhombic Bmab (Low Temperature Orthorhombic, LTO) [32], to pseudo-tetragonal Pccn [30] or P4$_2$/ncm (Low Temperature Tetragonal, LTO) [32], then to the HTT I4/mmm or $D_{4h}^{17}$, and occasionally to orthorhombic Fmmm (High-Stoichiometry Orthorhombic, HSO) [33] SG, depending on δ and T, as shown in Figure 1. Skinner *et al.* [33] reported the HSO phase at 150 °C with δ = 0.17, while Aguadero *et al.* [19] reported it for δ = 0.3. Biphasic regions have been evidenced as well [29,35,36]. These biphasic regions can be understood as the result of miscibility gaps which follow from the non-uniform mixing of interstitial oxygen defects into oxygen-poor and oxygen-rich domains. It is believed that strong interstitial oxygen correlations give rise to these miscibility gaps [29]. In these regions, the oxygen-poor domains take the maximum δ value of the bordering pure low- δ phase and the oxygen-rich domains the minimum δ of the pure high- δ phase. As shown in Figure 1, an excess oxygen concentration smaller than δ ≈ 0.01 is sufficient to induce phase separation into oxygen-poor and oxygen-rich phases. It should be noted that some of these transitions are indeed observed near room temperature (RT). Finally, the existence of a monoclinic (space group $C_2$) superstructure at $\delta = 0.25$ has also been reported at RT. This structure is four times larger than the average orthorhombic unit cell, resulting in the formation of a solid-solution with a composition of $La_8Ni_4O_{17}$ [37]. It will not be considered further in the following discussion as the $\delta$ values achieved in this study are below this extreme value.

From an experimental perspective, XRD analysis provides characteristic reflections that are useful for distinguishing between the different phases and in determining phase fractions in biphasic materials [31,36]. Due to the orthorhombic distortion, the LTO phase is characterized by the splitting of certain reflections with $h \neq k$, such as (200) / (020). In addition, weak superlattice reflections such as (212) or (032) are expected, an indication of coherent octahedral tilts. In the case of the LTT phase, the splitting of reflections with $h \neq k$ is absent, but the superlattice reflections (212) and (032) persist. This implies that mixed LTO/LTT phases can be identified by the coexistence of split and non-split reflections, along with the presence of two sets of the superlattice reflections for (212) or (032), due to the difference in lattice constants between the LTO and LTT phases. It is commonly accepted to give and compare lattice parameters on the basis of the $\sqrt{2}a$ x $\sqrt{2}$a x $c$ parameters of the parent HTT cell. Finally, the room temperature evolution of the lattice parameters as a function of $\delta$ is known from different works [29,31,36], which gave rather consistent values. In the composition range that correspond to the broad HTT phase field, one can estimate the chemically-induced strain then as follows: $\approx -5.10^{-3}$ (a-axis) and $\approx +8\ 10^{-3}$ (c-axis), with the consequence that the unit cell volume does not vary dramatically in this composition range, at least close to RT.

The Raman analysis of L2NO4 requires knowledge of the expected normal modes. The $k = 0$ normal-mode enumeration of the different expected phases is available in a number of studies [38–41]. As previously mentioned, the highest symmetry HTT phase adopts a tetragonal structure with space group I4/mmm or $D_{4h}^{17}$ (also non-conventional F4/mmm symbol is sometimes used for a face centered cell of the parent I4/mmm structure). The atomic displacement patterns corresponding to the different Raman modes of HTT tetragonal and Bmab orthorhombic L2NO4 are illustrated in Figure S1 and Figure S2, respectively, adapted from Ref. [15,42]. The HTT tetragonal primitive cell includes one formula unit and, as a result, there are 21 optical modes of $2A_{1g} + 2E_g + 3A_{2u} + B_{2u} + 4E_u$ symmetry, from which the 4 “gerade” modes are Raman-active. According to all available computations, these Raman-active modes consist of the out-of-plane (c-axis, $A_{1g}$) motion and the in-plane (a and b-axis, $E_g$) motion of O and La ions [15,42]. Close to the stoichiometric composition ($\delta \approx 0$), the RT (LTO) structure is the orthorhombic Bmab (or $D_{2h}^{18}$) space group with two formula units in the primitive unit cell. Here, the phase transition from the ideal tetragonal structure to the orthorhombic one is induced by the rotation of the rigid $NiO_6$ octahedra about the original [110] tetragonal axis [32]. The sense of rotation and displacement for neighboring Ni sites needs to be opposite in sign, which results in a doubled unit cell whose a and b axes are approximately 2 times larger than and 45° to the original a axis. As a result, 42 vibrational modes are expected ($5A_g$, + $3B_{1g}$ + $4B_{2g}$ + $6B_{3g}$ + $4A_u$, + $7B_{1u}$, + $8B_{2u}$, + $5B_{3u}$), in which the 18 gerade modes are Raman active.

As mentioned in [32] the LTO phase is also described with a slightly distorted orthorhombic cell, LTO, leading to the Pccn ($D_{2h}^{10}$) space group. In such a case, the irreducible representation becomes $8_{Ag} + 8B_{1g} + 10B_{2g} + 10B_{3g} + 11A_u + 11B_{1u} + 13B_{2u} + 13B_{3u}$. The 36 gerade modes are Raman active. Slightly increasing δ, and still close to RT, a second phase transition is observed leading to another tetragonal structure (LTT) ($P4_2$/ncm, $D_{4h}^{16}$). The structural difference between the LTO and LTT phases arise from the axis of rotation of the oxygen octahedral: in the LTT structure the $NiO_6$ octahedra rotate by equal amounts about the [110] and [1-10] axes of the simple tetragonal cell, doubling the in-plane unit cell as in the LTO case [32]. Again, the cell is doubled below the phase transition compared to the LTO one, which means that 81 modes ($5A_{1g} + 3A_{2g}, + 3B_{1g} + 5B_{2g}, + 10E_g, + 3A_{1u} + 8A_{2u} + 8B_{1u} + 3B_{2u} + 13E_u$) are expected. In this case, the $A_{2g}$, $A_{1u}$, $B_{1u}$ and $B_{2u}$ are silent so that 23 Raman-active modes are expected. Finally, the oxygen rich ($\delta \geq 0.15$) phase is sometimes indexed in the orthorhombic system (Fmmm, $D_{2h}^{23}$) or even F222, $D_2^7$), due to a slight orthorhombicity in the [110] direction of the I4/mmm structure [33]. In both cases, 21 modes are expected. For the Fmmm structure, the symmetry of the normal modes is the following: $2A_g + 2B_{2g} + 2B_{3g} + A_u + 4B_{1u} + 5B_{2u} + 5B_{3u}$, among them the gerade modes are Raman active. Intuitively, the Fmmm structure is expected to be only a small distortion of the I4/mmm phase. In this case, a splitting of the $E_g$ modes of the parent phase into modes of $B_{2g}$ and $B_{3g}$ symmetry is expected. For the less symmetrical F222 phase, the irreducible representations are $3A + 4B_1 + 7B_2 + 7B_3$. Among them, $3A + 3B_1 + 6B_2 + 6B_3$ are Raman (and IR) active. In the latter case, the inversion center in the unit cell is lost, meaning that the zone center vibrational modes are both Raman- and infrared- active.

In most cases, the knowledge the density of vibrational states (VDOS) is of fundamental importance, often a prerequisite for the study of disordered materials. The measured and calculated phonon dispersion curves of tetragonal L2NO4 are reported in [38,39,41]. From the reported cutoff frequencies, no first-order eigenmodes are expected above ~ 700 $cm^{-1}$. In the case of the HTT structure, this frequency corresponds to a $E_u$ symmetry zone center mode, actually observed at ~ 660 $cm^{-1}$.

Regarding L2NO4, the primary experimental Raman data are described in [15,27,28,43], while additional data for the isostructural cuprate material can be found in [11,16,28,44–46]. A comprehensive summary of these datasets is provided in Table S1. Notably, the lattice dynamics of the $La_2CuO_4$ orthorhombic phase at the center of the Brillouin zone has been extensively characterized in [42], which details both the vibrational frequencies of the normal modes and their corresponding eigenvectors.

However, research over the past decades has been predominantly focused on $La_2CuO_4$ -derived superconducting oxides, with most investigations targeting the structural and vibrational properties of these materials at temperatures significantly below room temperature (RT). Furthermore, as previously noted, discrepancies persist in the reported literature. These inconsistencies likely arise not only from experimental challenges but also from variations in sample nature and quality, contributing to a degree of scatter in the obtained results.

## 2. Experimental Methodology

$La_2NiO_{4+\delta}$ bulk samples were prepared using a modified Pechini method. The details of the preparation methods are described in [33]. $La_2NiO_{4+\delta}$ thin films were deposited by pulsed-injection metal organic chemical vapor deposition (PI-MOCVD) on Si substrates covered with a 100 nm Pt layer. The platinum layer ensures that no modes from the substrate are detected as the metallic layer reflects all the incident photons. Dense thin films of 40, 200 and 500 nm L2NO4 were deposited at 750 °C. The precursor solution was prepared by dissolving commercial (Strem chemicals) powders $La(tmhd)_3$ and $Ni(tmhd)_2$ (tmhd = 2,2,6,6-tetramethylheptane-3,5-dionate) in m-xylene solution (Alfa Aesar) with a La/Ni ratio of 3.15 and a solution concentration of 0.02 $mol{\cdot}l^{-1}$. The solution was injected into the reactor with a frequency of 4 Hz, opening time of 2 ms. The carrier gas concentration was fixed at 34% Ar/66% $O_2$ with a total pressure of 5 Torr. The thickness of the films was controlled by the number of pulses injected.

X-Ray Diffraction (XRD) in Bragg Brentano (θ-2θ) configuration were obtained using a Cu-Kα (λ = 1.54 Å) source on a Bruker D8 Advance diffractometer. The Raman spectra were recorded on two different Renishaw inVia Raman systems, mostly with an excitation wavelength at 532 nm. It was equipped with an air-cooled CCD detector, different specific gratings (1200, 1800 and 2400 grooves/mm), and dielectric rejection filters (cut-off frequency at about 80 $cm^{-1}$). Frequency calibration was performed using a silicon standard, whose frequency was set at 520.5 $cm^{-1}$. This configuration ensured a spectral resolution, more exactly an accuracy on the frequencies determination, much better than 1 $cm^{-1}$. Excitation wavelengths 488, 633 and 785 nm were periodically used to discriminate between Raman and photoluminescence (PL) origins, and to probe possible resonant conditions. Within the exception of samples close to stoichiometry, the line shape of the signals in the main wavenumber range investigated (≈ 80–1320 $cm^{-1}$) did not depend on the excitation wavelength, which in particular ensured that of the recorded signal were of Raman scattering origin. The *in situ* Raman measurements were conducted in a Linkam THMS600 temperature cell connected to different gas supplies, including a 10% $H_2$/Ar mixture, synthetic air and oxygen. Measurements were conducted in a backscattering configuration; a long working distance 50x optic was used to focus the incident light at the sample and collect the scattered radiation. As polycrystalline samples were investigated, the polarization of the scattered light was not analyzed.

## 3. Results

Initial measurements on L2NO4 thin films revealed that RS typically provided identifiable signatures with, however, some data variability, including broad signals and the presence of weak intensity peaks. Furthermore, similar observations were made when analyzing ceramic-type L2NO4 samples prepared via conventional and well-established synthesis methods. These findings highlighted the necessity of obtaining additional Raman data on this specific compound, particularly in its ceramic/polycrystalline form, to ensure the reliability of the study.

Consequently, we aimed to collect this information, which will be compared to existing studies and progressively presented in the following sections. In the subsequent step, we will focus on L2NO4 in its thin-film form.

The primary objective of this study was to identify and characterize the distinct Raman signatures corresponding to the various anticipated structural phases of the material. This involved analyzing the vibrational modes associated with each phase, thereby enabling a comprehensive understanding of their lattice dynamics and phase transitions. To this end, we first used a bulk sample prepared using the usual citrate route, then sintered to obtain a ceramic pellet. The bulk L2NO4 sample was successively heated for 9 hours at 575 °C in 5% $H_2$/Ar and for 1 hour at 200 °C in dry air, respectively, to obtain δ values close to 0 (reduced sample) and close to δ ~ 0.13 (oxidized sample). The corresponding RT diffraction patterns after the high temperature treatments are shown in Figure 2a-b. The XRD pattern of the reduced sample was effectively refined (Rietveld method) with an orthorhombic Bmab model with no additional peaks observed. Splitting of diffraction peaks such as (200) and (313) and appearance of Braggs peaks with $h + k = 2n + 1$ (where n is an integer) could be effectively observed, in complete agreement with the Bmab SG model, see for example [30,31], which confirms the LTO phase assignment. The refined cell parameters a = 5.5390 Å, b = 5.4669 Å and c = 12.5439 Å are in close agreement with published data for to δ ≈ 0 [29,31]. As expected, the SG of the oxidized sample was found to be tetragonal I4/mmm with refined cell parameters of a = 5.4674 Å (reduced cell parameter √2a) and c = 12.6691 Å, corresponding to δ ≈ 0.13, according to [29–31].

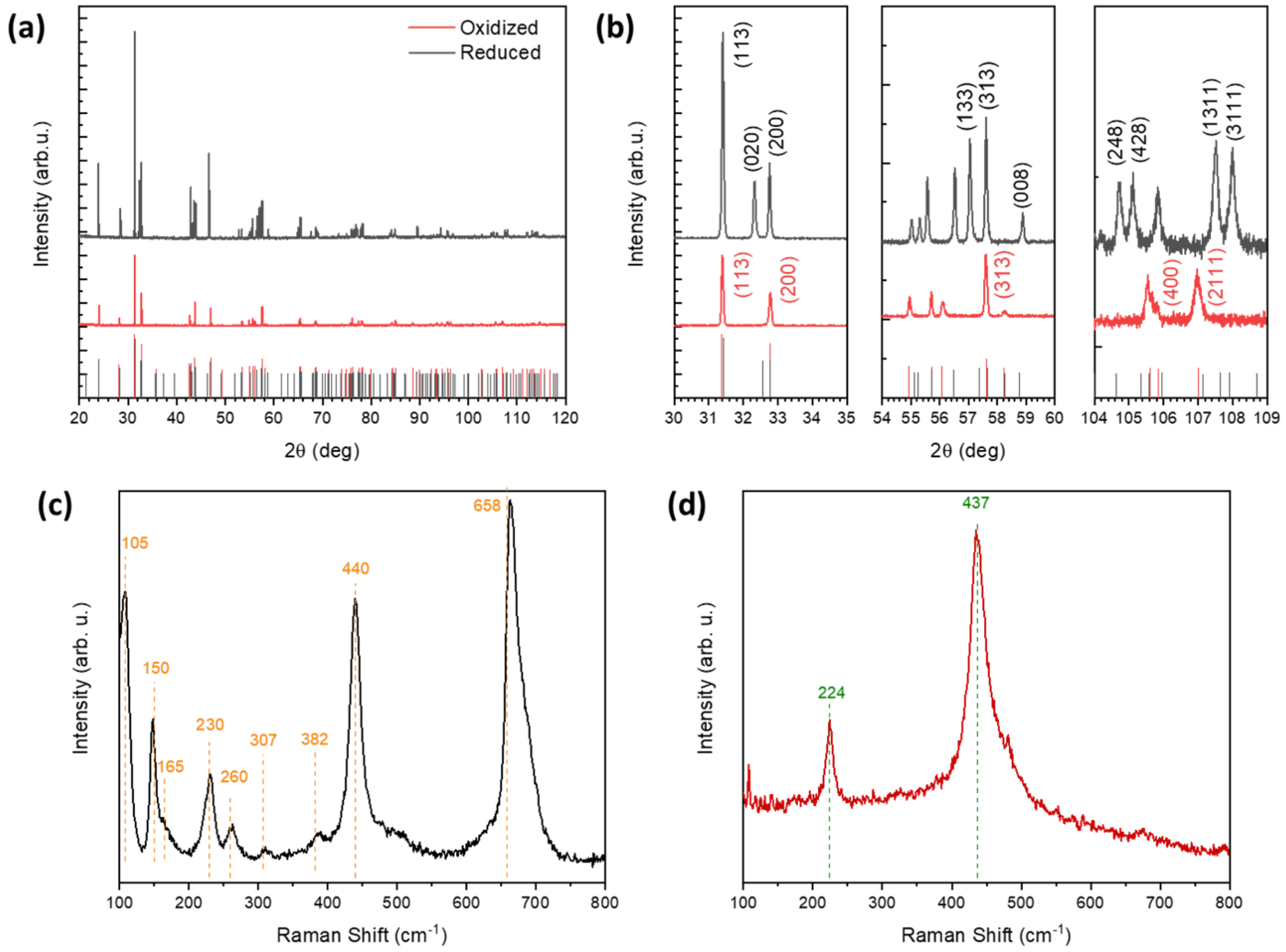

Figure 2: (a) XRD pattern of the reduced (stoichiometric, δ = 0) and oxidized sample (δ = 0.13), (b) magnified spectra showing the splitting of peaks in the LTO phase. The red markers correspond to the tetragonal I4/mmm phase (ICDD: 01-076-6116) and the black markers correspond to the orthorhombic Bmab phase (ICDD: 04-010-9709); Room temperature Raman spectra of the (c) reduced orthorhombic structure, and (d) oxidized tetragonal L2NO4 structure

Figure 2c-d shows the corresponding Raman spectra of the reduced and oxidized samples. As previously mentioned, as these L2NO4 samples are in the form of polycrystalline ceramics, selection rules cannot be considered. Nevertheless, it is possible to assign some of the peaks, as the frequencies and the corresponding ionic motions and Raman tensors are described in a number of previous studies [15,27]. As shown in Figure 2c and expected from the mode enumeration, many peaks were observed for the reduced sample (LTO phase), at about 105, 150, 165, 230, 260, 307, 382, 440 and 658 $cm^{-1}$. According to [42], other Raman-active modes are expected below the cutoff frequencies of the instruments used in the work. The lines observed at 224 and 675 $cm^{-1}$ appear as shoulders, but the resolution of the measurements is sufficient to evidence them. These lines were consistently observed for several oxidation/reduction cycles. As for the spectra obtained for the reduced phase, only the relative intensities fluctuate from one spectrum to the other.

Despite the limited data available in the literature for L2NO4, this spectrum looks similar to that given in references [27,47,48] with the exception of the doublet observed at around 658 $cm^{-1}$, which was not consistently observed throughout the references quoted above. This spectrum (Figure 2c) is also similar to those known for the isostructural cuprate materials [11,16,28,44–46], again with the exception of the high frequency 658 $cm^{-1}$ line, for the same reasons. From the polarized measurements conducted on single crystals [28], Raman modes at about 105, 225, 260, 440, and ≈ 500 $cm^{-1}$ are expected to be of $A_g$ symmetry. Without formal proof from our data, the other modes are expected to have $B_g$-type symmetry. According to the literature survey, no Raman-active mode is expected around 660 $cm^{-1}$. On the other hand, as already mentioned, this frequency corresponds quite closely to that of an active infrared mode of the HTT structure of $E_u$ symmetry with an observed low LO-TO (longitudinal optical and transverse optical) splitting, see for example [49].

For the oxidized sample (Figure 2d), two main lines are clearly observed, peaking at about 224 and 437 $cm^{-1}$, in agreement with most reports [15,27] and computation studies [42]. In the oxidized high δ stoichiometry state (HTT phase), 4 Raman-active modes are expected. The fully symmetric $A_{1g}$ modes are expected to have higher intensities than the in-plane $E_g$ modes. The symmetric $A_{1g}$ vibration of the apical oxygen atoms along the tetragonal axis is observed at about 440 $cm^{-1}$. In spite of some controversy, the second main mode at about 224 $cm^{-1}$ can be identified as the $A_{1g}$ vibration of the La atoms along the tetragonal c axis. We were able to further confirm this attribution by analyzing samples enriched in $^{18}O$. The in situ Raman spectra of the $^{18}O$ exchanged and completely back-exchanged (BE) L2NO4 thin film are shown in Figure S3. The peak of the $A_g$ mode, corresponding to the out-of-plane movement of the oxygen ion, shifts with changes in the 18O concentration, while the peak around 225 $cm^{-1}$ remains constant, which confirms that this mode is indeed not related to the movement of the oxygen ions but most probably the out-of-plane movement of the La ion.

In order to compare with literature data, we first probed the *in situ* LTT ($P4_2/ncm$) /LTO (Bmab) /HTT (I4/mmm) phase transition sequence as a function of temperature for δ close to 0. To this end, the sample was examined in a reducing atmosphere (Ar/5%H2), from liquid nitrogen temperature to about 550 °C. The evolution of the Raman spectra with increasing temperature is shown in Figure 3. The LTT/LTO phase transition is expected to occur in the -200 to -190 °C range [48,50,51], which according to Refs. [47,48] is a first-order phase transition. The spectra obtained at the lowest temperatures (-196 and -150 °C) are consistent with the RT spectrum discussed above. The main differences, depending on the observed modes, are some line shifts and strong differences in line widths. Only weak differences are observed comparing spectra obtained at -196°C and -150 °C, which mainly concern weak lines peaking below ca 300 $cm^{-1}$. This is in line with the only two references on this phase transition investigated using Raman spectroscopy. It seems that the temperature reached in the low-temperature cell was not low enough to fully observe the full phase transition [48]. Increasing the temperature, a strong broadening of all the lines is observed. Most of the modes that are characteristic of the LTO phase progressively shift and weaken, vanishing above 450 °C. In addition, there is a strong downshift of the low frequency mode observed at about 115 $cm^{-1}$ (104 $cm^{-1}$) at -196 °C and RT respectively. This mode is consistently assigned to the staggered rotational mode of $NiO_6$ octahedra, as described in Ref. [40,42] and effectively shows softening toward the orthorhombic-tetragonal phase transition temperature at 242 °C. It becomes overdamped above 350 °C, triggering the LTO / HTT phase transition. The behavior of this mode, in agreement with a second-order phase transition, closely follows the literature data that concern both L2NO4 and the isostructural cuprate compound [50,52].

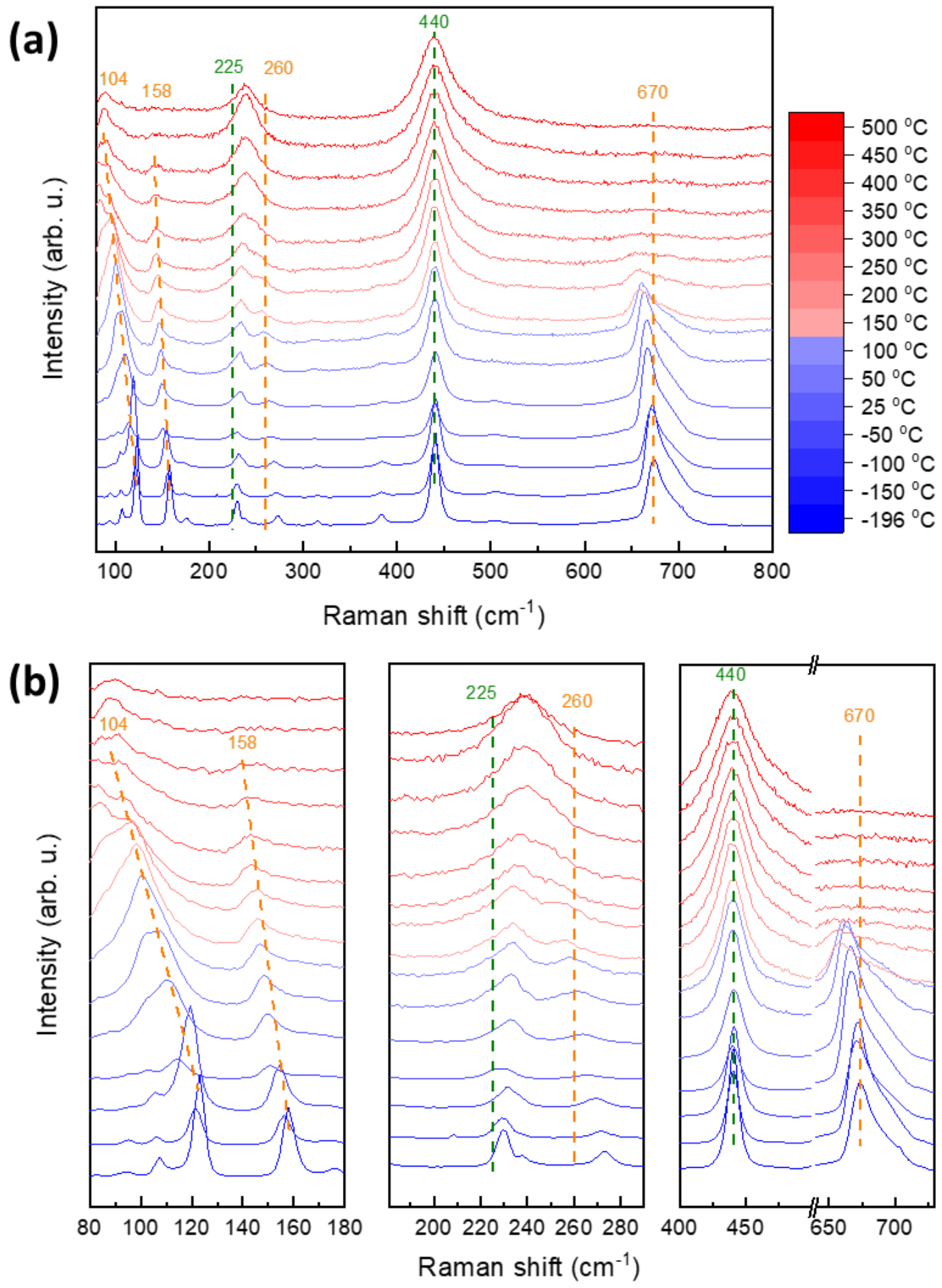

Figure 3: (a) *In situ* Raman spectra from -190 °C to 500 °C in dry air flow for bulk stoichiometry L2NO4 ($\delta = 0$) (b) magnified ranges to show evolution of different modes in selected regions

Of interest here are the frequencies of the lines peaking at about 223 and 437 $cm^{-1}$, that are common to the LTO and HHT phases. The linewidth of the line peaking at about 437 $cm^{-1}$ increases by a factor of about 3 in the temperature range investigated, while its frequency remains nominally constant to 1 or 2 $cm^{-1}$ within the measurement accuracy. While the line broadening seems consistent with the increase in temperature, this is clearly not the case with the rather subtle downshift observed. In most cases, line shifts of around 10 $cm^{-1}$ are expected in the temperature range investigated here [53]. This is also in line with most of the published results, see in particular [45]. The behavior of signals observed in the 200 to 300 $cm^{-1}$ frequency range will also be discussed. At low temperatures, 3 lines are clearly observed, which seem to merge into a single broad line for the highest temperatures. The apparent frequency of this line is 240 $cm^{-1}$, i.e., around 10 $cm^{-1}$ above the frequency expected for the HTT phase at high temperature. Based on the observations in Figure 3 , the data suggest that the LTO-HTT phase transition remains incomplete at 500°C.

In the second step, the reduced sample underwent gradual annealing in synthetic air to traverse the oxygen-doping phase diagram as slowly as possible. This process was conducted within a temperature range below 250°C, where all previously described phases, along with their respective two-phase coexistence regions, are expected to be observed. For each temperature, multiple spectra were recorded until there were no further observable changes in the line shape. At these temperatures the oxygen incorporation kinetics is expected to be slow enough to be able to record subtle changes in the line-shape of the spectra vs exposition times. Below 200 °C, the spectra remained invariant, as shown in Figure 4a. At 200 °C, we found a progressive disappearance of the characteristic signals of the LTO phase (around 105, 260 and 660 $cm^{-1}$) with time in favor of those of the HTT phase (only two peaks around 225 and 440 $cm^{-1}$), as indicated by the Raman spectra in Figure 4b-c. This gradual phase transition was observed *in situ* after around 1500 s. Notably, this transition is evidenced by the apparent average downshift of spectral lines from 240 $cm^{-1}$ to 225 $cm^{-1}$, and the broadening of the 440 $cm^{-1}$ peak. Once again, no clear or systematic shift was observed in the spectral line common to all phases, specifically the vibrational mode at around 440 $cm^{-1}$, as a function of oxygen enrichment. This implies that the so-called chemical expansion cannot be unambiguously detected in this specific system. This stands in contrast to observations in certain non-stoichiometric oxides, where such effects are more clearly measurable, and also deviates from the initial expectations of this study. Finally, note the sharp decrease in overall signal intensity during this transition, which accompanies the change in color of the sample from yellowish to black (see Figure 4d-e).

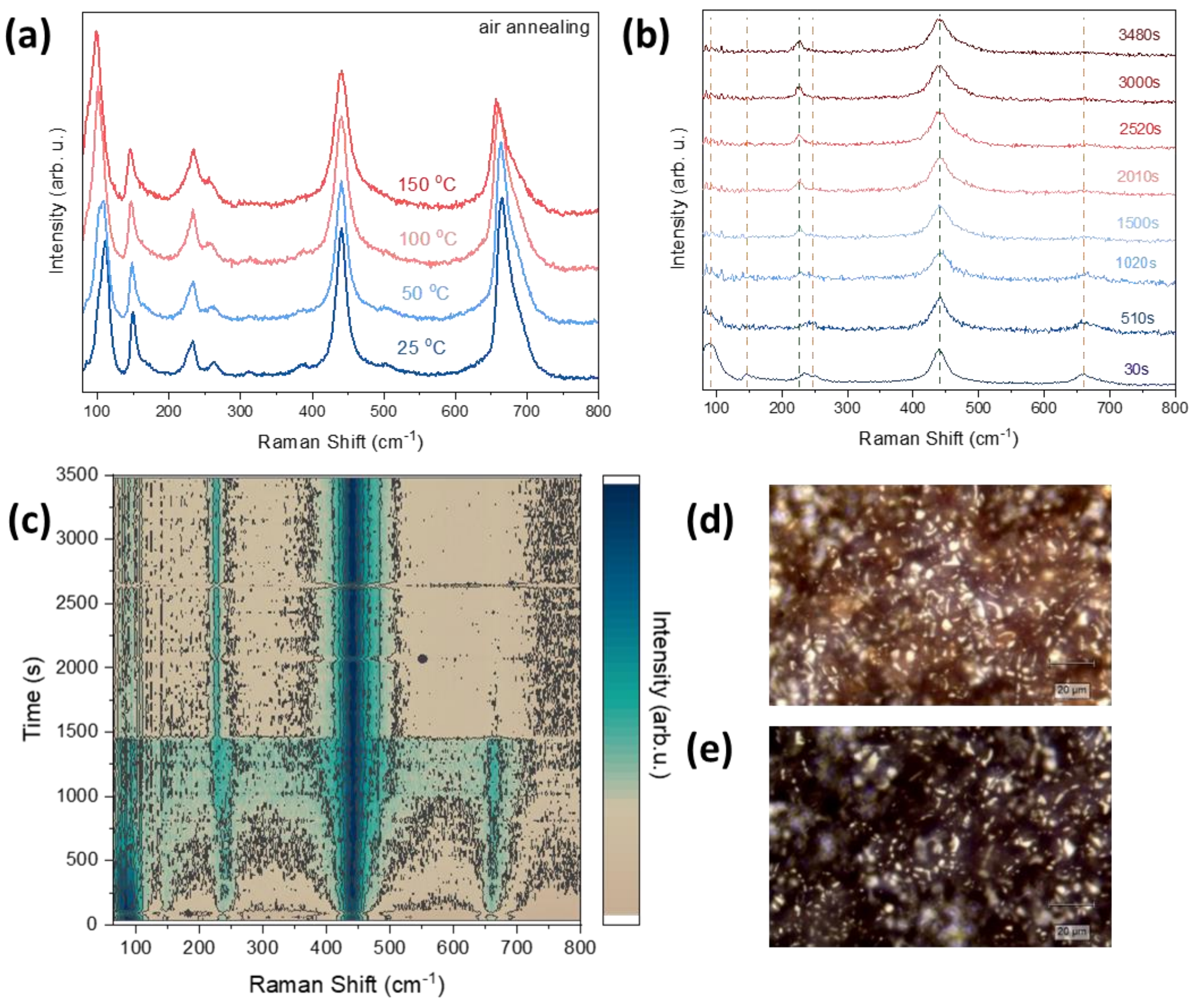

Figure 4: *In situ* evolution of the Raman spectra of stoichiometric bulk L2NO4 in dry air at (a) temperatures from 25 to 150 °C, (b-c) at 200 °C with time. The positions of the peaks corresponding to the L2NO4 modes from tetragonal (green dotted lines) and orthorhombic (orange dotted lines) phases are marked in the graph; (d-e) Optical surface images of the bulk pellet showing the color evolution from (d) reduced to (e) oxidized phase L2NO4 at 200 °C

Similar oxidation/reduction *in situ* protocols were repeated for L2NO4 dense 200 nm thin films deposited by MOCVD. Optimization of deposition conditions to obtain a La/Ni ratio as close as possible to 2 is described elsewhere [25]. In all cases, the diffraction patterns could be indexed according to the HTT tetragonal phase. We initially encountered significant variations in the Raman signals, particularly as a function of film thickness, *c.f.* Figure 5. As demonstrated in the following analysis, we successfully identified the two distinct Raman modes characteristic of the HTT phase. However, these Raman modes were superimposed on a broad background signal, along with weaker spectral features, in particular in the low frequency range of the spectra. In general, the thinner the layer, the more intense this broad signal becomes, as illustrated in Figure 5. This broad signal has an apparent maximum at around 575 $cm^{-1}$, and a cut-off frequency at around 750 $cm^{-1}$. Notably, this cut-off frequency aligns with known experimental and computed vibrational state density of states. It is likely that this signal partly reflects this oxygen partial VDOS, influenced by the small grain size of the crystals (10–20 nm) forming the layers and the high proportion of grain boundaries associated with them (see Figure S4 for SEM and XRD of the 500 nm thin film). Such a high defect density in those grain boundary regions may effectively induce a partial relaxation of the k ≈ 0 selection rule. From this point onward, films with a thickness of 200 nm or greater were used for subsequent studies as the spectra show a smaller number of peaks arising from defect associated with the small grain size of thin films.

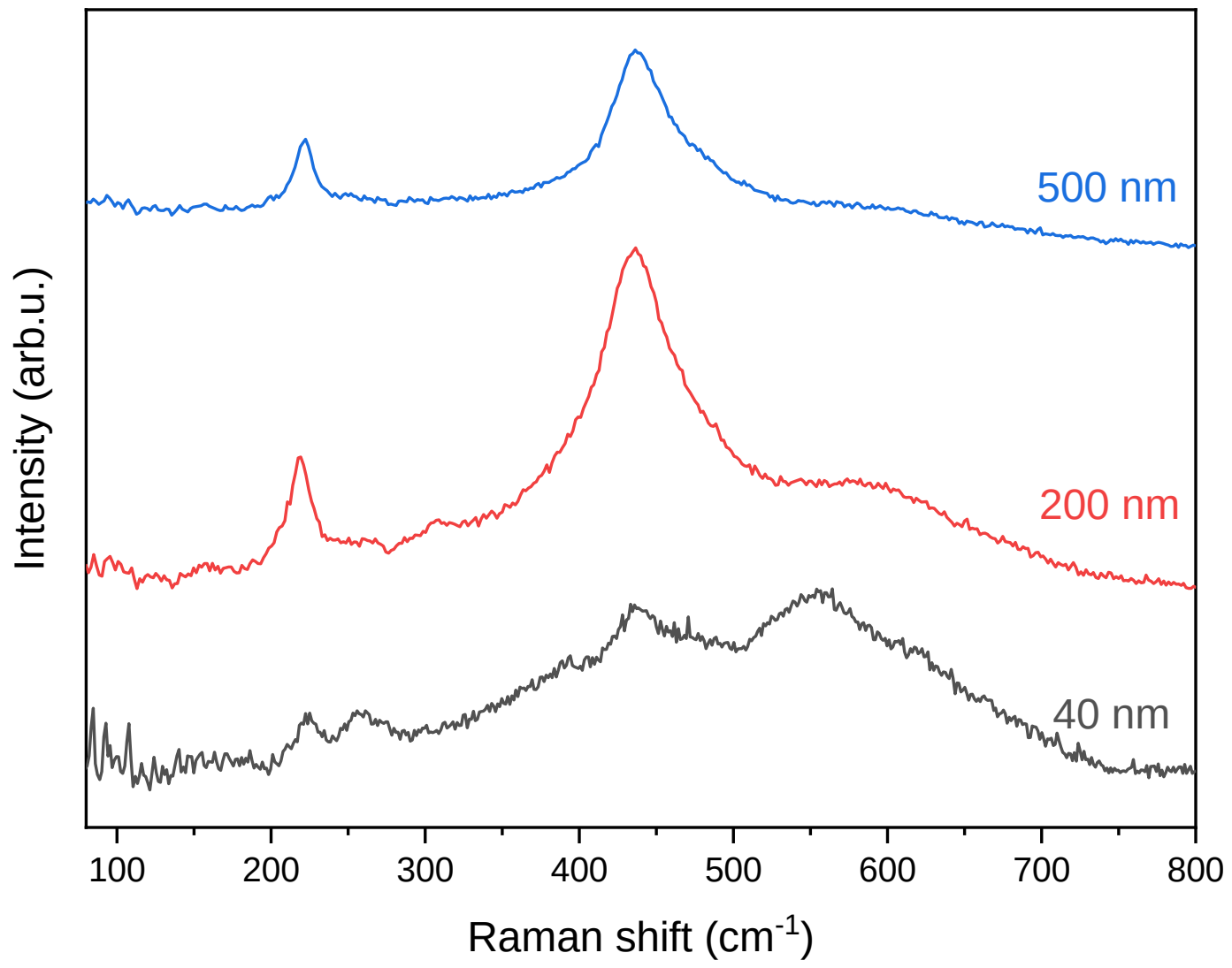

Figure 5: Raman spectra of L2NO4 thin films with thicknesses of 40, 200 and 500 nm

The *in situ* oxidation tests yielded results similar to those of the bulk material: while the LTO phase was observed, the Raman signals evolved rapidly over time—even at temperatures below 200°C—making it nearly impossible to monitor the intermediate phases, as shown in Figure 6a. The *in situ* reduction of the thin films produced interesting results. Surprisingly, the film got reduced in less than 300 seconds at 50 °C, as shown in Figure 6b. The typical LTO modes appear, including the soft mode around 99 $cm^{-1}$ which is associated with HTT to LTO phase transition. The film was measured at different points, confirming a homogenous reduction across its entire surface. For comparison, the reference bulk sample reduced to the LTO phase at a much higher temperature (450 °C), whereas the thin film exhibited remarkably fast reduction kinetics at a very low temperature (50 °C). There was no further change in the line spectra for 1800 s, so the temperature was increased to 100 °C where the only change observed was the increase in the peak width and intensity of the mode at 660 $cm^{-1}$.

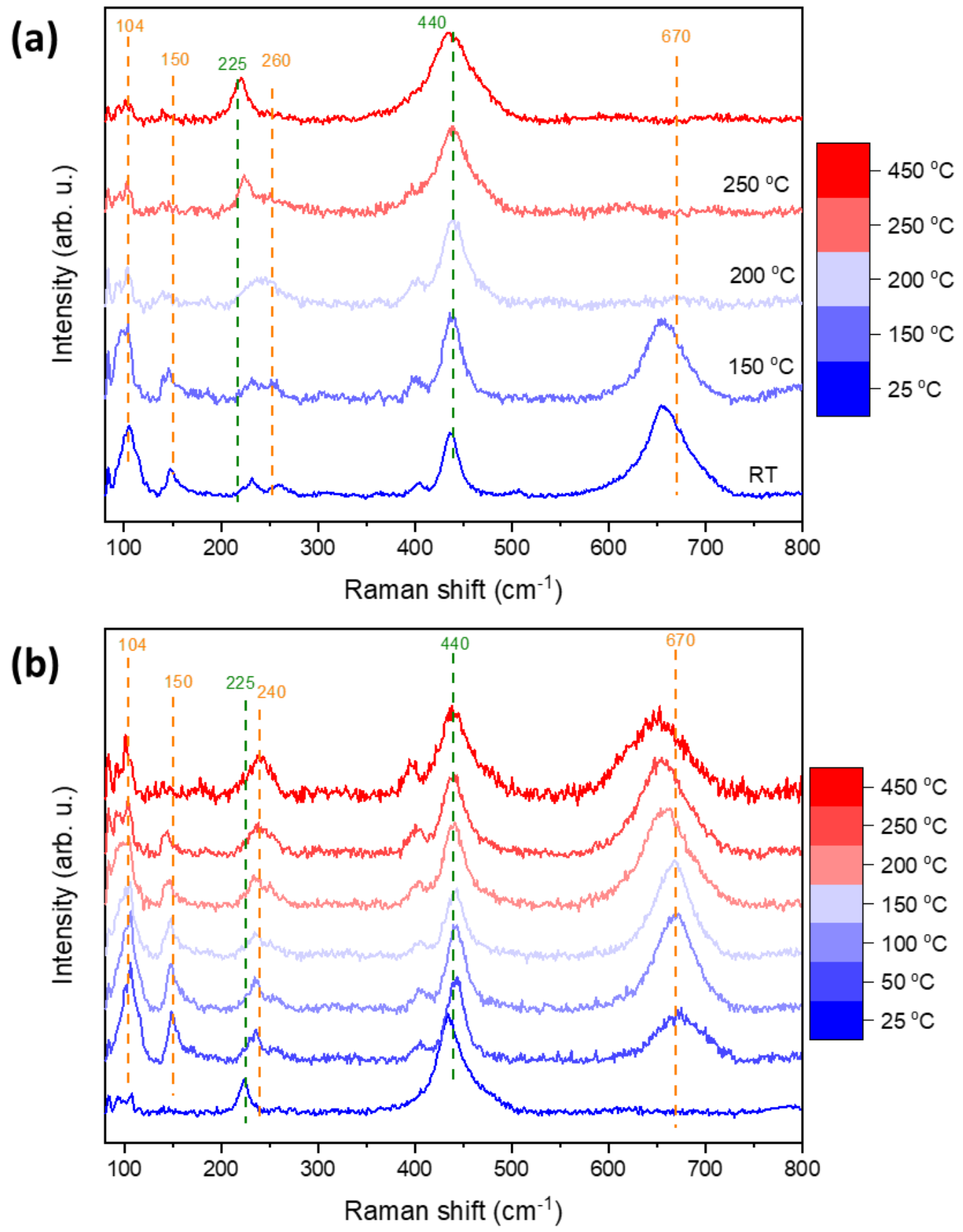

Figure 6: *In situ* Raman spectra of a L2NO4 thin film annealed at different temperatures in: (a) oxidizing conditions; (b) reducing conditions

## 4. Discussion

Our findings underscore the challenges inherent in working with ultrathin layers, particularly when the material exhibits nanoscale grain sizes (approximately 10 nm) and a high density of grain boundaries. From the standpoint of the techniques employed in this study, the primary effect of using thin films instead of bulk samples was a pronounced broadening of all spectral signals, accompanied by additional features induced by the high density of crystalline defects—most notably, the broad Raman band centered at approximately 550 $cm^{-1}$.

Our results prompt several questions. First, can we conclusively attribute the observed Raman signatures to the three anticipated phases? The HTT phase Raman signature is effectively the expected one, irrespective of material processing (thin film or bulk). The LTO phase Raman signature was most likely obtained for bulk L2NO4, insofar the XRD pattern revealed the expected lattice parameters and confirmed the orthorhombic distortion. Moreover, the observed Raman signature closely aligns with those reported in previous studies. We successfully detected comparable Raman signals for materials in the form of thin films; however, some degree of variability was observed in the measurements. Regarding the LTT phase, the measurements were not entirely conclusive, indicating potential uncertainties or limitations in the data obtained. Based on the analysis of all the proposed phase diagrams, theoretically this specific phase should have been observed, particularly during the annealing of the ceramic material in synthetic air at a relatively low temperature (200 °C). However, the absence of this phase in the results suggests a discrepancy with the expected behavior.

Moreover, a striking finding was the high to very high intensity of the line (in fact a doublet) centered at about 660 $cm^{-1}$ systematically observed close to $\delta \approx 0$ for all the samples studied in both in the ceramic or thin film form. According to all computations, Raman-active modes are not expected to appear within this frequency range. Different hypothesis are to be considered: i) parasitic phases. None of the signatures related to the parasitic phases likely to be observed, whether due to the synthesis process, sample storage, or even wrong analysis conditions can account for the observed signal. ii) Aliovalent doping on the lanthanum sites. The literature survey clearly shows that such a substitution, when exceeding a few atomic percent, can induce significant modifications in the line-shape of the spectra. For instance, studies on the isostructural $La_2CuO_4$ compound, as shown in references [46,54], provide examples of these effects. In such a case, symmetry-breaking arguments are commonly invoked, allowing for the detection of symmetry-forbidden Raman lines. Nevertheless, such an argument does not apply in our case, as samples from two distinct synthesis processes exhibited the same trends. Therefore, there is no rationale for expecting the doping effects to be similar across both synthesis methods. iii) specific resonant conditions. Why should we consider this hypothesis? Resonant scattering has indeed been observed for the isostructural $La_2CuO_4$ compound close to the exact stoichiometry, tuning the excitation wavelength in the near infrared range (1064 nm) [55]. For this specific situation, a strong resonant one-phonon scattering was observed for seven peak frequencies that correspond to longitudinal optic (LO) phonons with both $E_u$ and $A_{2u}$ symmetry. The results were interpreted on the basis of a Fröhlich-type electron-phonon interaction mechanism.

For all the samples examined in this study, particularly near $\delta = 0$, the spectra show a clear dependence on the excitation wavelength. This is illustrated in Figure S5, which pertains to the thin film of L2NO4. We focus on this sample for several reasons: i) the peak around 660 $cm^{-1}$ is not observable at 785 nm, but becomes detectable at approximately 633 nm, becoming clearly visible at wavelengths below 532 nm. This specific mode is effectively resonantly enhanced. The frequency of this mode corresponds closely to that of an $E_u$ mode in the HTT parent phase, making the hypothesis of a Fröhlich-type interaction plausible. In this situation, multiple harmonics of such a mode are often observed. The first harmonic of this mode has been clearly observed. Concerning L2NO4, only one IR-active mode should be concerned by this specific interaction. ii) Figure 4 shows that this peak around 660 $cm^{-1}$ vanishes at the onset of oxidation. This may explain why this line has not been observed systematically, as it is only detectable very close to $\delta = 0$. Without precise control of the hyper-stoichiometry, it becomes challenging to observe this line consistently. iii) Apart from this specific mode, all other modes become clearly observable beyond 532 nm, which supports the same conclusion. However, it should be noted that the electron-phonon interaction to be considered is different for the Raman-allowed modes. Finally, the wavelength dependence of the Raman spectrum was also examined for the oxidized sample, that concerns the tetragonal phase. No specific wavelength dependence of the spectra was observed in this case as shown in Figure S5.

Regarding the line shape of the broad signal observed around 575 $cm^{-1}$ in thin films it can be correlated to the small average size of the crystallites that constitute the film. This line does not strongly depend on the excitation wavelength, exhibiting the same cut-off frequency at about 750 $cm^{-1}$ and must therefore be correlated with disorder considerations.

## 5. Conclusions

This work was devoted to a systematic study of the Raman spectra of $La_2NiO_{4+\delta}$ samples in the form of bulk pellets and thin films. To eliminate any signal arising from grain size effects and/or defects typically present in thin films, the reference spectra of the reduced orthorhombic phase (LTO) and the oxidized tetragonal phase (HTT) were obtained using a $La_2NiO_{4+\delta}$ sintered pellet free of impurities. The sharp peak around 660 $cm^{-1}$ in the reduced samples (both bulk and thin films) which had been previously correlated with forbidden infrared mode was concluded to be due a specific resonance effect. As for the thin films, a broad Raman peak observed around 575 $cm^{-1}$ in oxidized samples could be correlated to the small grain size (~20 nm). Moreover, phase transitions from LTO ($\delta \approx 0$) to HTT ($\delta > 0.1$) in both the pellet and thin films, as well as the reverse transition, were successfully observed *in situ* using Raman spectroscopy. However, due to fast oxygen exchange kinetics it was not possible to obtain the Raman spectra for the expected intermediate phases. *In situ* Raman studies also revealed a fast reduction of the $La_2NiO_{4+\delta}$ thin films at a very low temperature of 50 °C in a reducing atmosphere of 10% $H_2$/Ar. Finally, it was not possible to extract $\delta$ values from the frequencies of the different Raman lines, which was one of the information sought at the start of this work.

## Supporting Information

All sample datasets and materials related to this work are made available under CC BY 4.0 license in the zenodo repository: 10.5281/zenodo.15209655

## Acknowledgements

This project has received funding from the European Union's Horizon 2020 research and innovation program under grant agreements no. 824072 (Harvestore project) and no. 101017709 (EPISTORE project). This research has benefited from characterization equipment of the Grenoble INP - CMTC platform supported by the Centre of Excellence of Multifunctional Architectured Materials "CEMAM" n°ANR-10-LABX- 44-01 funded by the "Investments for the Future" Program. The authors acknowledge Prof. Stephen J. Skinner (Imperial College London, U.K.) for preparing and providing the reference $La_2NiO_4$ pellets.

## Supplementary Information

Table S1: Calculated and observed Raman mode positions and mode assignment for L2NO4 and $La_2CuO_4$

| **Tetragonal I4/mmm** | | | **Orthorhombic Bmab** | | | | | |
|---|---|---|---|---|---|---|---|---|
| **Calculated** | | **Observed (Pellet)** | **$La_2CuO_4$** | | **L2NO4 Observed** | **Mode Assignment** | | |
| L2NO4<br>[15] | $La_2CuO_4$<br>[15] | L2NO4<br>[27] | Calculated<br>[42] | Observed<br>(Single Crystal)<br>[28] | Bmab<br>[27] | Orthorhombic (Bmab)<br>[27,42] | | |
| | | | | | | | | |
| $E_g$<br>83 | $E_g$<br>79 | - | $B_{1g}$<br>116 | | 63 | $A_g(1)$ | $O_{eq}$<br>bending | |
| $A_{1g}$<br>138 | $A_{1g}$<br>141 | $A_{1g}$ (La)<br>180-200 | $B_{3g}$<br>117 | | 87 | $A_g(4)$,<br>$B_{3g}(5)$<br>$A_g(5)$,<br>$B_{3g}(6)$ | $O_{ap}$<br>La<br>bending | |
| $E_g$<br>252 | $E_g$<br>246 | $E_g$ ($O_{ap}$)<br>220 | $A_g$<br>124 | $A_g$<br>126 | 109 | $B_{2g}(4)$<br>$B_{1g}(3)$ | La<br>bending | |
| $A_{1g}$<br>450 | $A_{1g}$<br>431 | $A_{1g}$ ($O_{ap}$)<br>437 | $B_{2g}$<br>143 | | 151 | $A_g(3)$ | La<br>stretching | |
| | | $A_g$ ($O_i$)<br>322 | $B_{3g}$<br>155 | | 169 | $B_{3g}(4)$ | La<br>stretching | |
| | | | $A_g$<br>166 | $A_g$<br>156 | 235 | $B_{2g}(3)$ | $O_{ap}$<br>bending | |
| | | | $A_g$<br>217 | $A_g$<br>229 | 263 | $B_{1g}(2)$ | $O_{ap}$<br>bending | |
| | | | $B_{1g}$<br>225 | | 335 | $B_{3g}(2)$<br>$B_{2g}(2)$ | Ni- $O_{eq}$<br>bending | |
| | | | $B_{1g}$<br>240 | | 387 | | $O_{ap}$<br>stretching | |
| | | | $B_{3g}$<br>242 | | 443 | $A_g(2)$<br>$B_{3g}(3)$ | $O_{ap}$<br>stretching | |

| | | | | | | | |
|---|---|---|---|---|---|---|---|
| | | | $B_{2g}$ 269 | | 685 | $B_{3g}(1)$ $B_{2g}(1)$ | Ni- $O_{eq}$ Stretching |
| | | | $B_{3g}$ 328 | | | | |
| | | | $A_g$ 328 | $A_g$ 273 | | | |
| | | | $B_{2g}$ 410 | | | | |
| | | | $A_g$ 417 | $A_g$ 426 | | | |
| | | | $B_{3g}$ 482 | | | | |
| | | | $B_{1g}$ 579 | | | | |
| | | | $B_{3g}$ 723 | | | | |

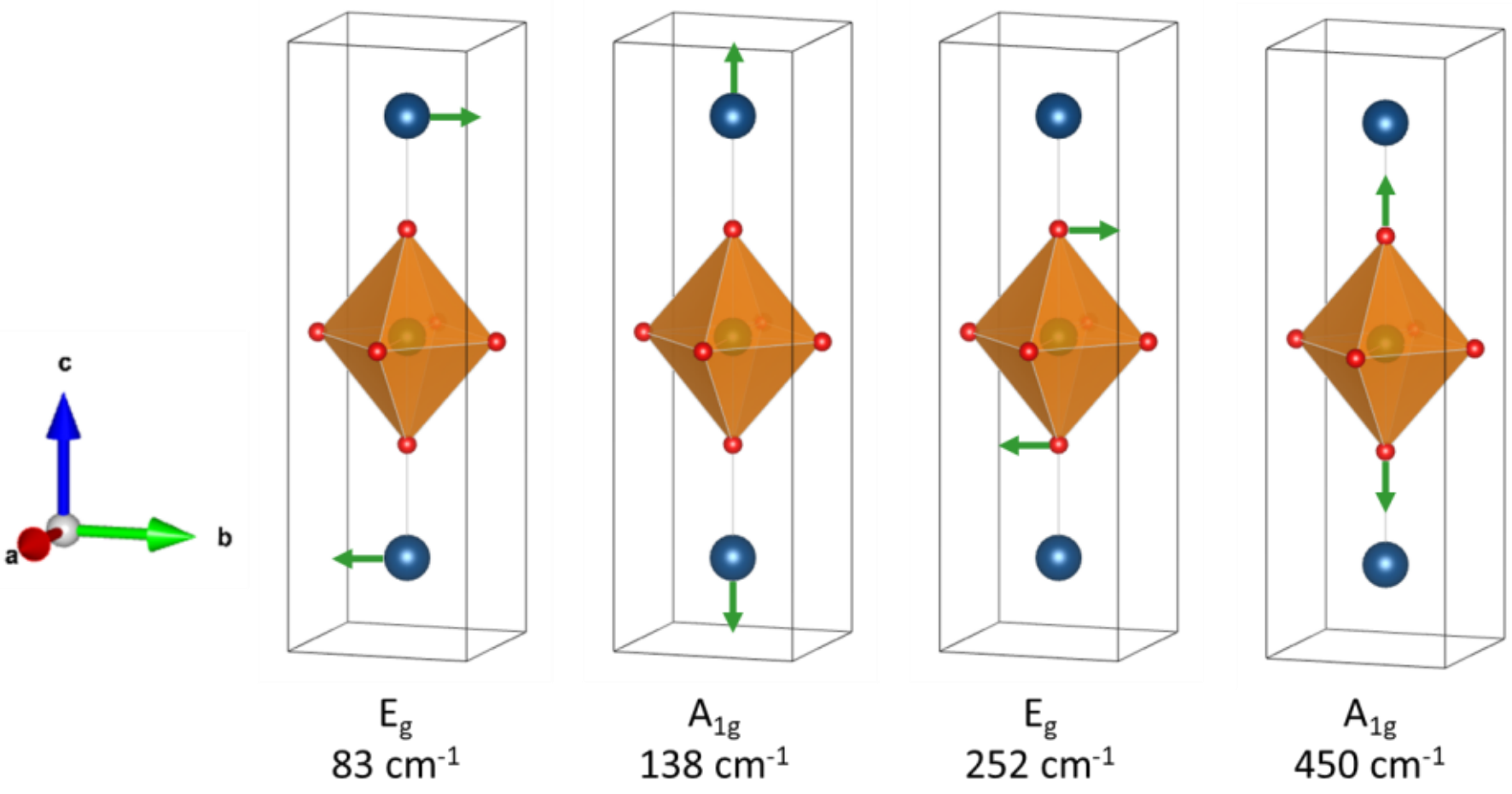

Figure S1: Calculated and assigned Raman modes of Tetragonal (SG: I4/mmm) L2NO4 [15]

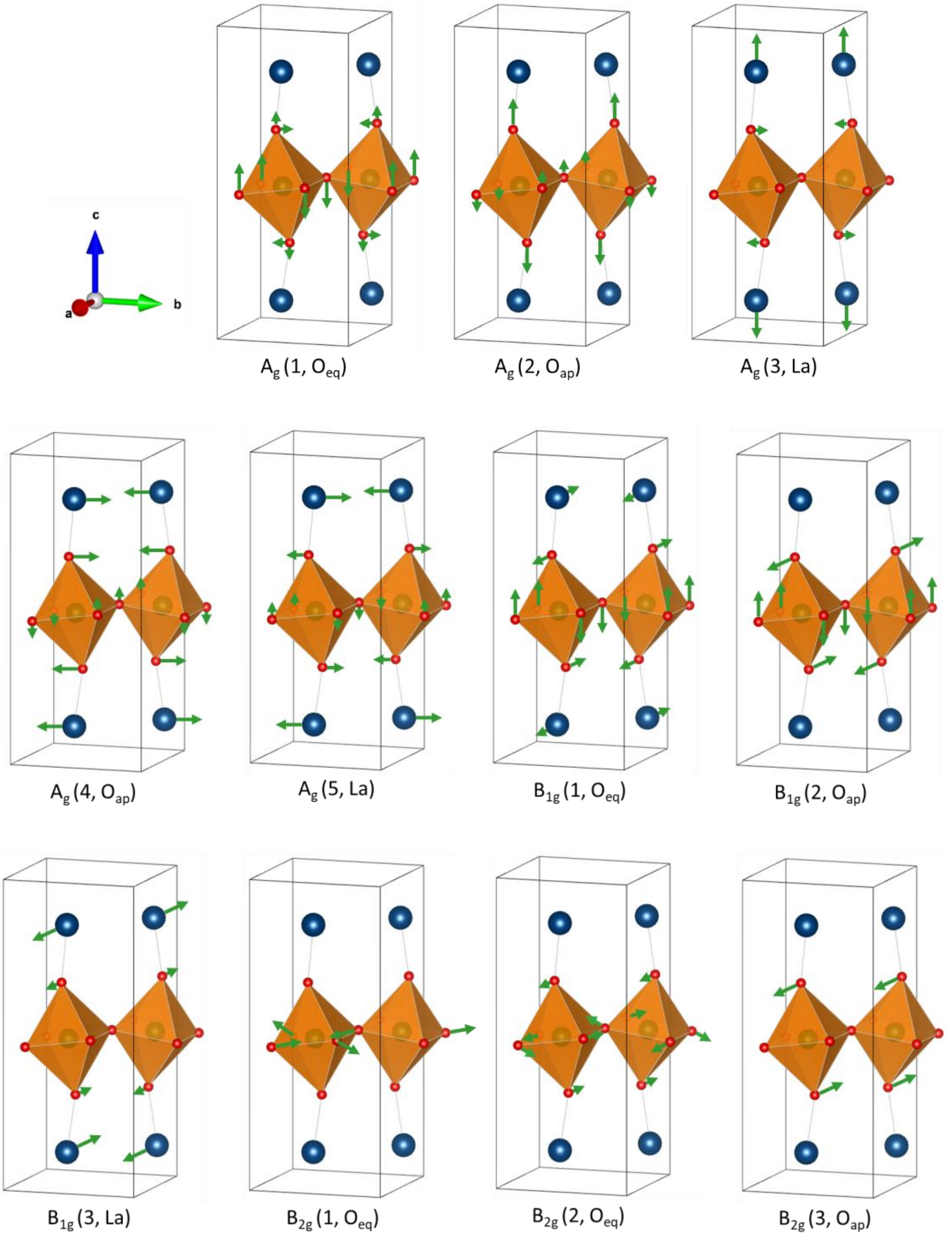

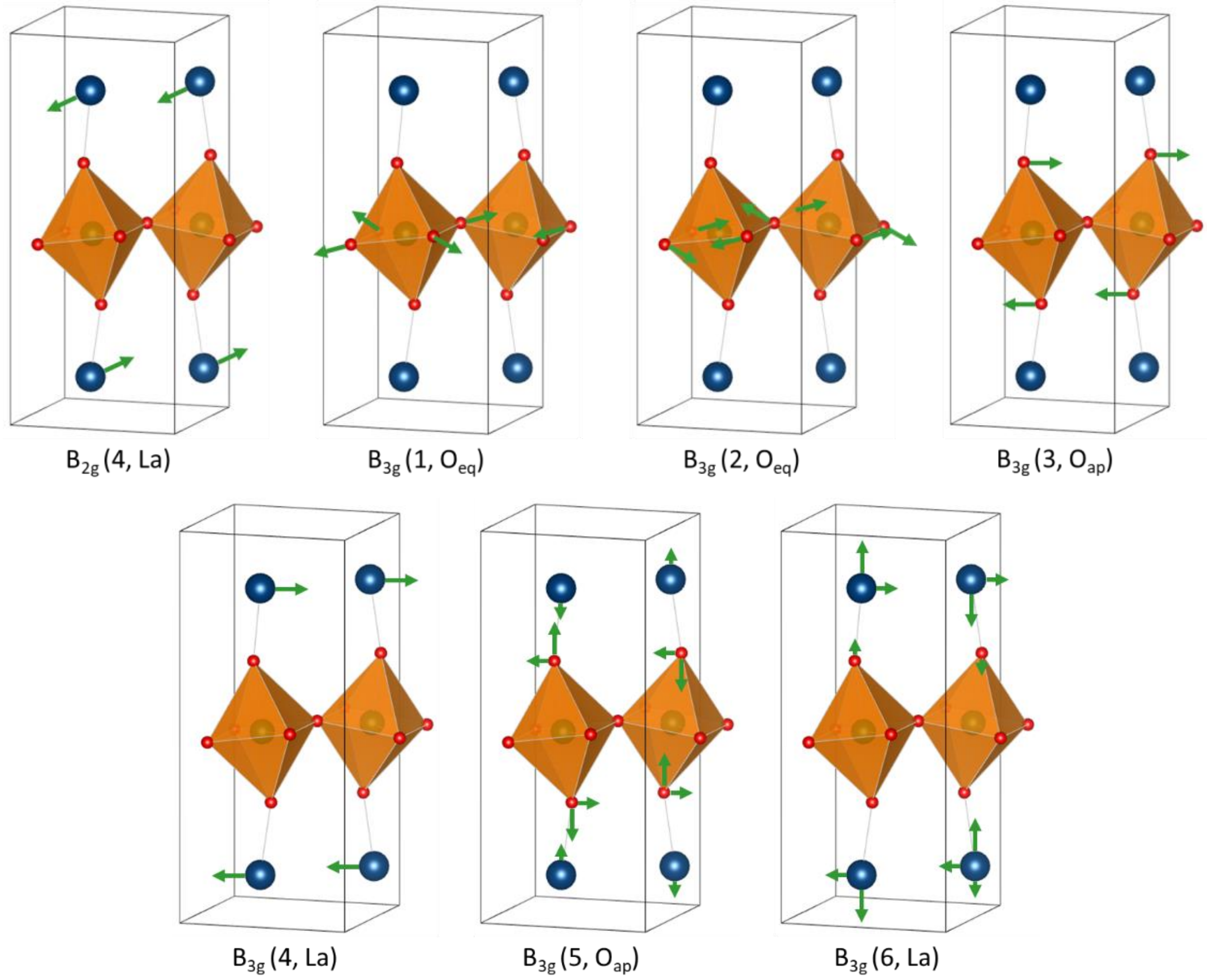

Figure S2: Raman modes of stoichiometric L2NO4 with orthorhombic phase (SG: Bmab). The length of the arrow indicates the amplitude (large, intermediate and small) – amplitudes are not scaled [42]

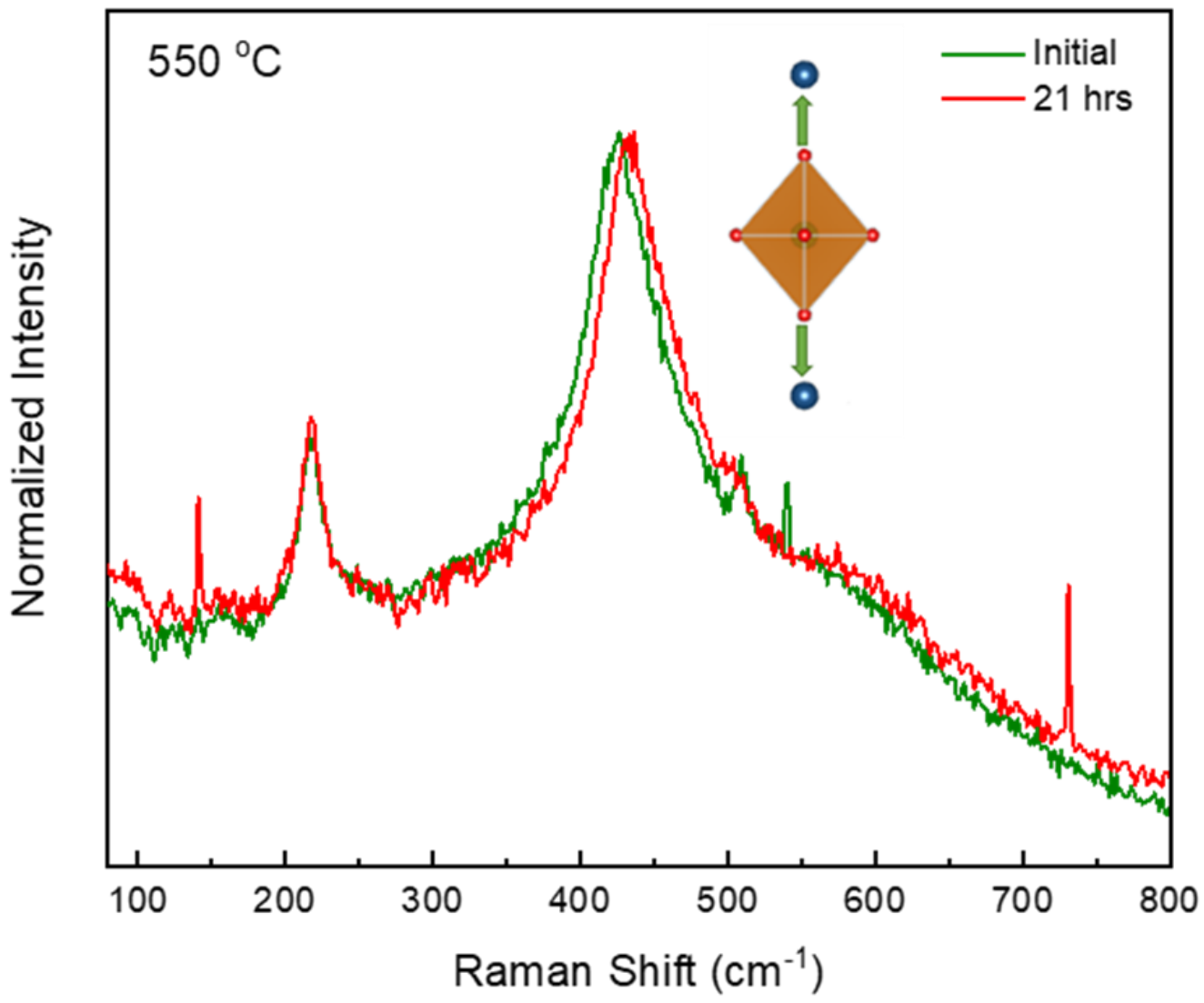

Figure S3: Raman spectra of the $^{18}O$ exchanged film (initial, green line) and the isotopic shift of the oxygen mode observed during the back-exchange (21 hrs, red line) at 550 °C of 200 nm L2NO4 thin film

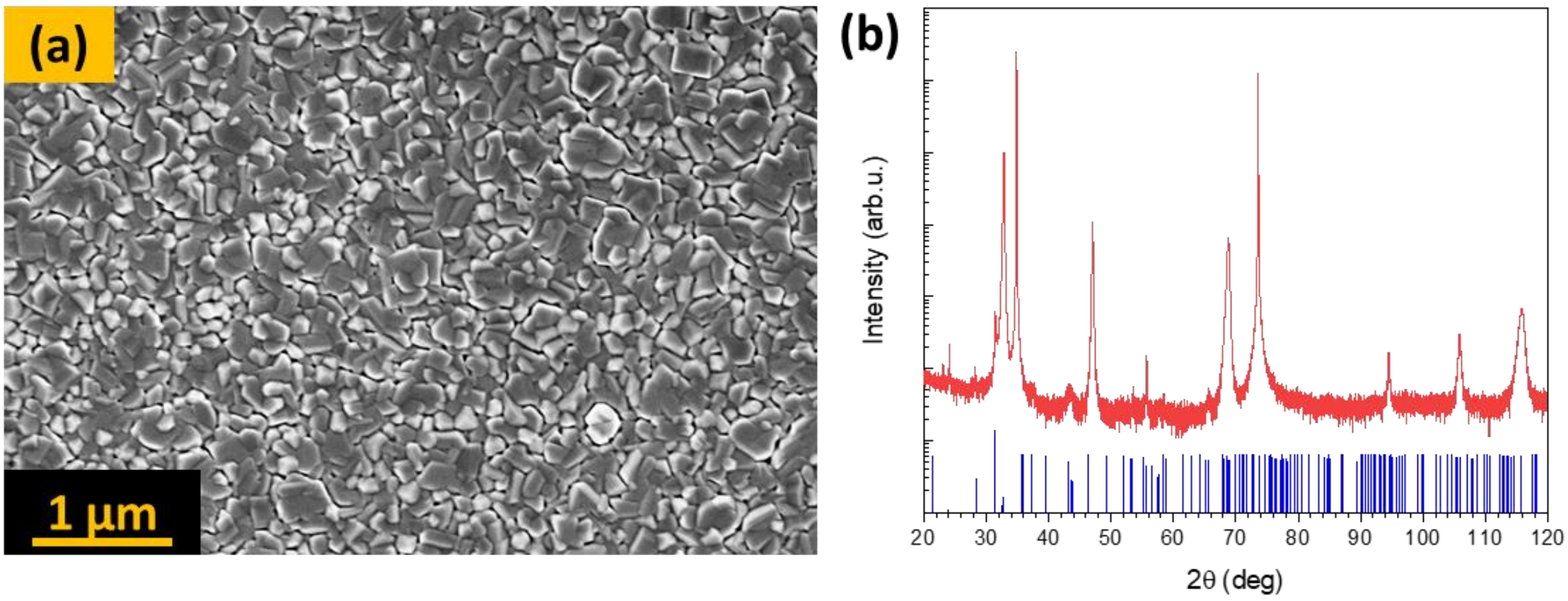

Figure S4: (a) SEM image showing the surface microstructure, and (b) XRD of the 500 nm L2NO4 thin film

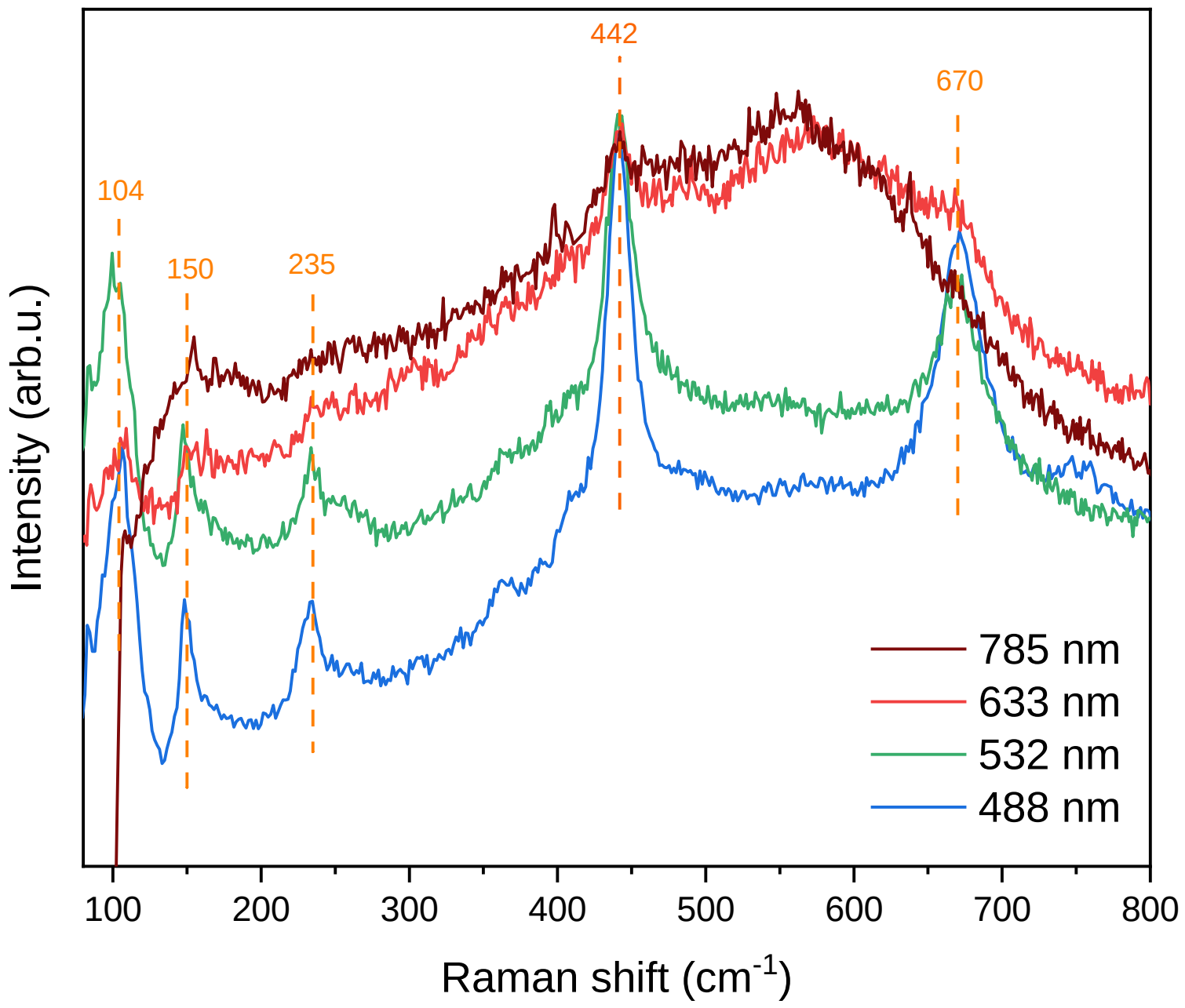

Figure S5: Raman spectra of reduced L2NO4 thin film (δ = 0) measured with different excitation wavelength of 488, 532, 633 and 785 nm.